\documentclass[10pt, conference]{IEEEtran}
\usepackage{cite}
\usepackage{amsmath,amssymb,amsfonts}
\usepackage{algorithm}
\usepackage{algpseudocode}
\usepackage{graphicx}
\usepackage{textcomp}
\usepackage{xcolor}
\usepackage{tabularx}
\usepackage{booktabs}
\usepackage{bm}
\usepackage{comment}
\usepackage{braket}
\usepackage[T1]{fontenc}
\usepackage{notoccite}
\usepackage{subcaption}
\def\BibTeX{{\rm B\kern-.05em{\sc i\kern-.025em b}\kern-.08em
    T\kern-.1667em\lower.7ex\hbox{E}\kern-.125emX}}
\usepackage{url}
\usepackage{multirow}
\begin{document}

\title{
Practical Quantum Federated Learning for Privacy-Sensitive Healthcare: Communication Efficiency and Noise Resilience
\thanks{}}

\author{Suzukaze Kamei\IEEEauthorrefmark{1}, 
Hideaki Kawaguchi\IEEEauthorrefmark{1},
and Takahiko Satoh\IEEEauthorrefmark{2} \\
[1ex]
\IEEEauthorrefmark{1}\textit{Graduate School of Science and Technology, Keio University, Yokohama, Kanagawa 223-8522 Japan} \\
\IEEEauthorrefmark{2}\textit{Faculty of Science and Technology, Keio University, Yokohama, Kanagawa 223-8522 Japan} \\
\{suzukaze\_kamei, hikawaguchi, satoh\}@keio.jp
}

\maketitle

\begin{abstract}

AI-driven medical diagnostics increasingly requires collaborative model training across institutions, yet centralizing patient data conflicts with privacy regulations. Federated Learning enables distributed training without raw data sharing, but remains vulnerable to gradient inversion and model leakage attacks. Furthermore, harvest-now-decrypt-later attacks render computationally secure protocols insufficient for protecting long-lived medical records. Quantum communication offers information-theoretic security immune to such threats, making Quantum Federated Learning (QFL) a compelling framework for healthcare. However, practical deployment is constrained by communication overhead and quantum channel noise. We present a systematic quantitative study of communication, convergence, and noise trade-offs in QFL, introducing two complementary strategies to reduce quantum transmissions: (1) structured parameter reduction via light-cone feature selection in parameterized quantum circuits, and (2) a Hybrid QFL architecture that dynamically switches between centralized and decentralized aggregation. We show that Hybrid QFL reduces total quantum transmissions from $3\,TNMP$, the cost of pure Centralized QFL, to $\{3t + 2(T - t)\}\,NMP$ over $T$ rounds while preserving near-centralized convergence. We further demonstrate that decentralized aggregation is more noise-resilient under depolarizing noise, and evaluate Steane code-based quantum error correction in high-noise regimes. Our results provide an integrated design framework for communication-efficient, noise-aware QFL, clarifying practical trade-offs for scalable quantum-secure distributed learning in healthcare.

\end{abstract}


\section{Introduction}
\label{sec:introduction}

The rapid expansion of AI-driven medical diagnostics has created an urgent demand for collaborative model training across hospitals and clinical institutions. However, centralizing patient records to train shared models conflicts with privacy regulations 
such as HIPAA~\cite{hipaa1996} and GDPR~\cite{gdpr2016}, making direct data sharing infeasible in practice. Federated Learning (FL)~\cite{mcmahan2017communication} addresses this need by enabling collaborative model training without centralizing raw data, allowing institutions to retain local control over sensitive records.

However, recent studies have shown that gradient inversion and model reconstruction attacks can extract sensitive information from shared gradients~\cite{zhu2019deep}, revealing that classical FL does not fully guarantee the privacy of shared parameters. Existing countermeasures such as differential privacy~\cite{geyer2018differentiallyprivatefederatedlearning, mcmahan2018learningdifferentiallyprivaterecurrent} and homomorphic encryption~\cite{aono2017privacy} improve security but introduce performance degradation or computational overhead. More fundamentally, these approaches rely on computational hardness assumptions. This makes them vulnerable to harvest-now-decrypt-later (HNDL) attacks, in which adversaries collect encrypted gradients today and decrypt them once sufficient computational power becomes available, a particularly acute threat for long-lived medical records whose sensitivity persists for decades.

\begin{figure*}[htbp]
    \centering
    \includegraphics[width=1\linewidth]{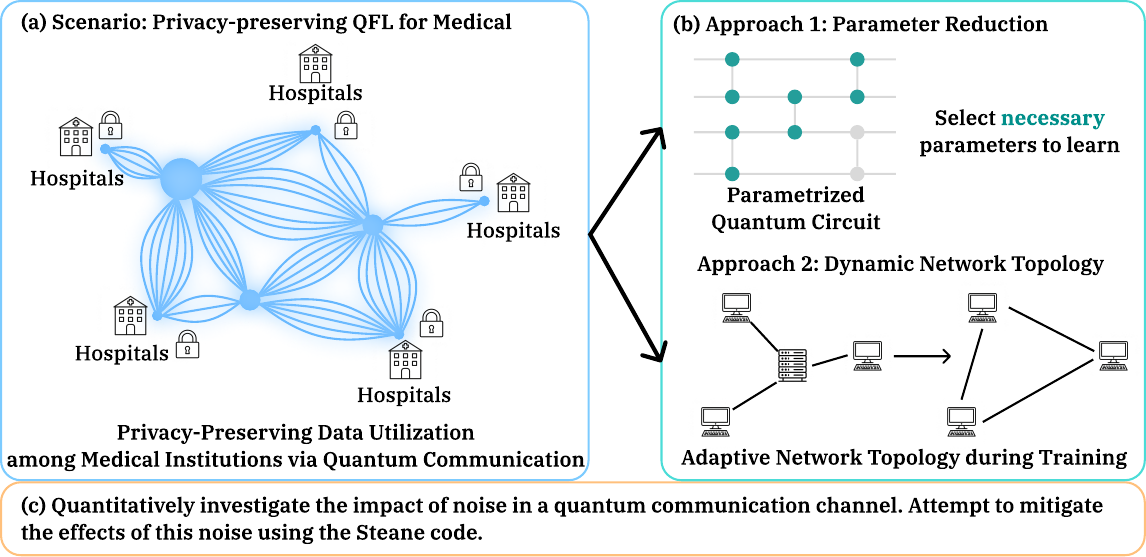}
    \caption{Overview of this work: (a) Quantum Federated Learning utilizing medical data via quantum communications. (b) Two approaches are proposed to reduce the number of quantum transmissions in Federated Learning: the first approach reduces the parameter count by selecting and aggregating only the necessary parameters. The second approach utilizes a Dynamic Network Topology to modify the topology during the training process. (c) Furthermore, we investigate the impact of inherent noise in current quantum communication channels and evaluate whether applying the Steane code can mitigate these noise effects.}
    \label{fig:overview_of_work}
\end{figure*}

Quantum communication offers a fundamentally different security paradigm. Unlike classical cryptography, quantum communication enables information-theoretic protection of model parameters during aggregation, e.g., via Greenberger-Horne-Zeilinger (GHZ)-based Quantum Secure Aggregation (QSA)~\cite{zhang2023federatedlearningquantumsecure}, whose security holds unconditionally regardless of an adversary's computational resources. Quantum Federated Learning (QFL) combines FL with quantum communication for the aggregation phase, thereby mitigating classical-channel leakage and hindering direct gradient reconstruction. However, practical implementation faces two major obstacles:

\begin{enumerate}
    \item Quantum communication overhead, which scales with the number of clients, 
    parameters, and training rounds.
    \item Quantum channel noise, which accumulates across repeated transmissions and 
    can destabilize learning.
\end{enumerate}

Furthermore, since gradient inversion attacks become more effective as the amount of shared parameter information increases~\cite{wu2023learning, geiping2020inverting}, minimizing transmitted information is not only a communication issue but also a security requirement.

Despite growing interest in QFL, a systematic study of the trade-offs between communication cost, convergence dynamics, and noise resilience remains lacking. This work addresses these gaps not by proposing fundamentally new quantum protocols, but by providing a systematic design and analysis framework that clarifies the practical trade-offs among communication cost, convergence, and noise resilience in QFL. In particular, three fundamental questions remain open:

\begin{itemize}
    \item Circuit-Communication Scaling: Can circuit-level causal structures be exploited to reduce the number of transmitted parameters?
    \item Topology Adaptation: How does dynamic aggregation topology affect communication efficiency and convergence stability?
    \item Noise Resilience \& Mitigation: How does quantum channel noise impact foundational QFL architectures, and can error correction effectively mitigate it?
\end{itemize}

This work addresses these questions through an integrated design and analysis framework, as illustrated in Fig.~\ref{fig:overview_of_work}.

First, to study circuit-communication scaling, we propose Approach 1: Parameter Reduction utilizing light-cone feature selection in parametrized quantum circuits (Fig.~\ref{fig:overview_of_work}b upper side). By exploiting light-cone within Quantum Neural Networks (QNNs), we reduce the number of aggregated parameters while preserving model expressivity.

Second, to investigate the impact of network structure on communication efficiency, we propose Approach 2: Dynamic Network Topology (Fig.~\ref{fig:overview_of_work}b lower side). We introduce a hybrid architecture that dynamically transitions from centralized to decentralized QFL during the training process. This allows us to reduce overall transmission frequency while maintaining convergence stability.

Third, to quantify the trade-offs between noise resilience and communication cost, we investigate the impact of inherent depolarizing noise in quantum communication channels on foundational QFL architectures (Fig.~\ref{fig:overview_of_work}c). Furthermore, we evaluate Steane code-based quantum error correction~\cite{steane1996error} to assess the trade-off between noise mitigation and physical qubit overhead.

The main contributions of this work are as follows:

\begin{enumerate}
    \item Quantitative integration of light-cone feature selection into QFL (Approach 1). Building on the light-cone feature selection framework proposed by Suzuki et al.~\cite{suzuki2025light}, we systematicall integrate this mechanism into the QFL aggregation phase and quantitatively demonstrate its effect on quantum transmission reduction, establishing an explicit link between circuit causality and communication efficiency in federated quantum settings.
    \item Dynamic topology for communication efficiency (Approach 2). 
    We introduce a hybrid aggregation architecture that adapts its topology during training, showing that transmission frequency can be significantly reduced without destabilizing convergence.
    \item Noise mitigation under realistic quantum channels. 
    We quantify the impact of depolarizing noise on foundational QFL architectures and evaluate Steane code-based error correction, clarifying the trade-offs between robustness and physical qubit overhead.
    \item Unified communication cost framework for QFL. We establish a unified formulation of quantum communication cost across centralized, decentralized, and hybrid QFL architectures, deriving explicit scaling laws as functions of $N$, $M$, $P$, and $T$. This framework enables systematic comparison of architectural trade-offs and provides a quantitative basis for scalability analysis in practical QFL deployment.
\end{enumerate}

The remainder of this paper is organized as follows: Section~\ref{sec:relatedwork} reviews related previous studies, and Section~\ref{sec:proposedmethods} details the core technologies utilized in our methodology. Section~\ref{sec:experimentalsetup} describes the experimental setup of this study, followed by the results and discussion in Section~\ref{sec:resultsanddiscussion}. Finally, Section~\ref{sec:conclusion} concludes the paper.

\section{Related Work}
\label{sec:relatedwork}

We review prior work on FL taxonomy, quantum secure aggregation, feature selection in quantum circuits, and non-IID data.

\subsection{Classification of FL}

First, the classification of FL is explained. Here, classifications based on data distribution and network structure are discussed.

\subsubsection{Classification by Data Distribution}

FL can be categorized into three types based on data distribution characteristics~\cite{yang2019federated, ballester2025quantum}:

\begin{itemize}
    \item Horizontal FL: Datasets share the same feature space but differ in samples. A typical example is cross-device learning, such as Google's Android update framework~\cite{mcmahan2017communication}.
    
    \item Vertical FL: Datasets share overlapping samples but differ in feature spaces, e.g., different departments within the same hospital contributing complementary attributes of the same patients.
    
    \item Federated Transfer Learning: Both samples and feature spaces differ across participants, requiring transfer learning techniques to align representations.
\end{itemize}

We focus on the horizontal FL setting, which is the most widely adopted and communication-intensive scenario.

\subsubsection{Classification by Network Structure}

Based on the network topology, existing FL frameworks can be broadly categorized into Centralized FL, Decentralized FL, and Hierarchical FL~\cite{liu2020client, briggs2020federated}. In Centralized FL, a central server coordinates the learning process by aggregating local updates and distributing the global model. In contrast, Decentralized FL relies on a peer-to-peer architecture where clients collectively perform model aggregation through local communication, eliminating the need for a central server. Furthermore, Hierarchical FL introduces intermediate edge servers to perform partial aggregation, reducing the communication overhead between clients and the core cloud.

However, these conventional architectures are inherently static throughout the training process. This static nature often forces a persistent trade-off: centralized topologies provide stable convergence but suffer from communication bottlenecks, while decentralized topologies alleviate these bottlenecks but can exhibit slower convergence. Therefore, dynamically adapting the network topology during training to leverage the advantages of both architectures remains an important but underexplored challenge.

\subsection{Security Threats and the Case for Information-Theoretic Protection}

Beyond gradient inversion attacks that operate on transmitted updates~\cite{zhu2019deep, wu2023learning, geiping2020inverting}, FL in healthcare faces a more fundamental long-term threat: the HNDL attack. In this attack model, an adversary passively records encrypted model updates during training and stores them until sufficient computational power becomes available to break the underlying encryption, whether through advances in classical cryptanalysis or the advent of cryptographically relevant quantum computers~\cite{mosca2018cybersecurity}. This threat is particularly acute for medical data such as genomic records, whose sensitivity persists for decades and whose exposure could have irreversible consequences for patients and their families.

Patient records carry legally and ethically enforced confidentiality obligations that extend across decades; a gradient exchanged during FL training today may expose sensitive health information long after the training process concludes. Standard countermeasures, differential privacy and homomorphic encryption, rely on computational hardness assumptions that HNDL attacks are specifically designed to circumvent.

Information-theoretic security, achieved through quantum communication protocols such as QSA, provides protection that holds unconditionally, independent of an adversary's current or future computational resources~\cite{zhang2023federatedlearningquantumsecure}. This motivates the adoption of QFL as the appropriate framework for privacy-critical healthcare applications, and distinguishes it from classical secure aggregation approaches.

\subsection{Quantum Aggregation Protocol}
\label{sec:qsa_protocol}

We adopt a GHZ-state-based QSA protocol~\cite{zhang2023federatedlearningquantumsecure}, assuming a semi-honest model for both the server and participating clients. A detailed description is provided in Section~\ref{sec:proposedmethods}. Here, we summarize the essential mechanism.

In centralized FL, the server prepares a GHZ state with one qubit per client and distributes them accordingly. Each client encodes its model parameters onto the received qubit via an $R_Z$ rotation and returns it to the server. By performing inverse operations and measurement, the server obtains the aggregated sum of the parameters without accessing individual contributions. In decentralized settings, one client assumes the aggregation role, while the aggregation principle remains the same.

Regarding security, the protocol employs decoy-state techniques to detect external eavesdropping, where an attack can be detected with probability $1 - (3/4)^d$ using $d$ decoy states~\cite{zhang2023federatedlearningquantumsecure}. The protocol is analyzed under a semi-honest assumption for internal participants, ensuring that individual model parameters cannot be inferred from the shared GHZ state and that only aggregated information is accessible. Detailed proofs are provided in the original work.

In this study, we focus on improving communication efficiency and robustness while adopting the established security guarantees of the QSA protocol.

\subsection{Feature Selection in Quantum Circuits}

We briefly summarize the feature selection framework based on quantum kernels proposed in~\cite{suzuki2025light}. The specific implementation adopted in this study is described in Section~\ref{sec:proposedmethods}.

In this approach, a quantum circuit is used to obtain the quantum kernel. The circuit contains rotation parameters $\theta$ and auxiliary coefficients (denoted $\lambda$ in~\cite{suzuki2025light}) that account for contributions from partial traces over unmeasured qubits. The method evaluates the contribution of circuit parameters to the predictive performance and selects only the most influential features.

In the context of this study, we leverage this feature selection mechanism to reduce the number of effective trainable parameters, thereby decreasing the amount of information transmitted during quantum aggregation.

\subsection{Independent and Identically Distributed (IID) and Non-IID Data}

In FL settings, data distributions vary across clients. IID settings assume that each client possesses statistically equivalent datasets. In practical applications, however, data are typically non-IID due to heterogeneity in user behavior, device characteristics, or institutional differences.

Non-IID data are known to degrade convergence speed and model performance in FL~\cite{zhao2018federated}. Since communication-efficient training is particularly challenging under non-IID settings, evaluating robustness in such scenarios is essential for practical deployment.

\begin{figure}[htbp]
    \centering
    \includegraphics[width=\linewidth]{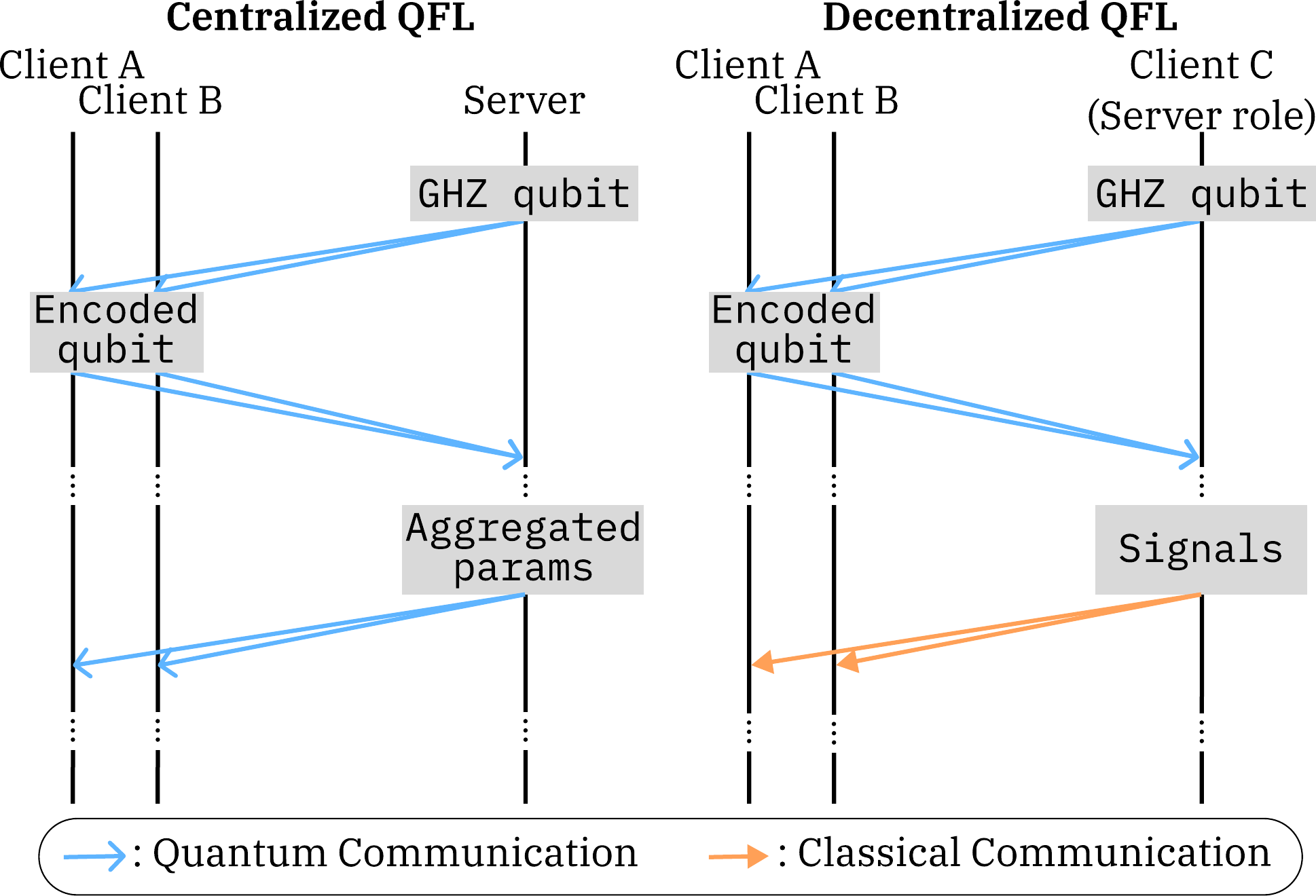}
    \caption{Centralized QFL and Decentralized QFL Protocols: Both protocols begin with GHZ-state generation and parameter encoding. They differ in the redistribution phase. Centralized QFL redistributes the updated global model via quantum teleportation, whereas Decentralized QFL rotates the aggregation role among clients without redistributing a global model.}
    \label{fig:centralized_decentralized}
\end{figure}

\section{Proposed Methods}
\label{sec:proposedmethods}

We define QFL as an FL framework in which parameter aggregation and transmission are performed using quantum communication protocols. Unlike classical FL, QFL leverages QSA and quantum teleportation to transmit model parameters while maintaining information-theoretic security. It should be noted that the proposed QFL maintains classical model training, while incorporating quantum communication in the aggregation phase.

\subsection{Centralized QFL and Decentralized QFL}

Building on Section~\ref{sec:relatedwork}, we describe the concrete implementations of Centralized and Decentralized QFL in our framework.

The protocol flows of Centralized and Decentralized QFL are shown in Fig.~\ref{fig:centralized_decentralized}. In Centralized QFL, model parameters are aggregated between the server and clients using the secure aggregation protocol. After averaging, the updated global model is redistributed via quantum teleportation between the server and each client. In Decentralized QFL, clients alternately assume the aggregation role. The acting client aggregates parameters from others, updates its model locally, and notifies the next acting client via classical communication.

\subsection{Hybrid QFL}

\begin{figure}[htbp]
    \centering
    \includegraphics[width=\linewidth]{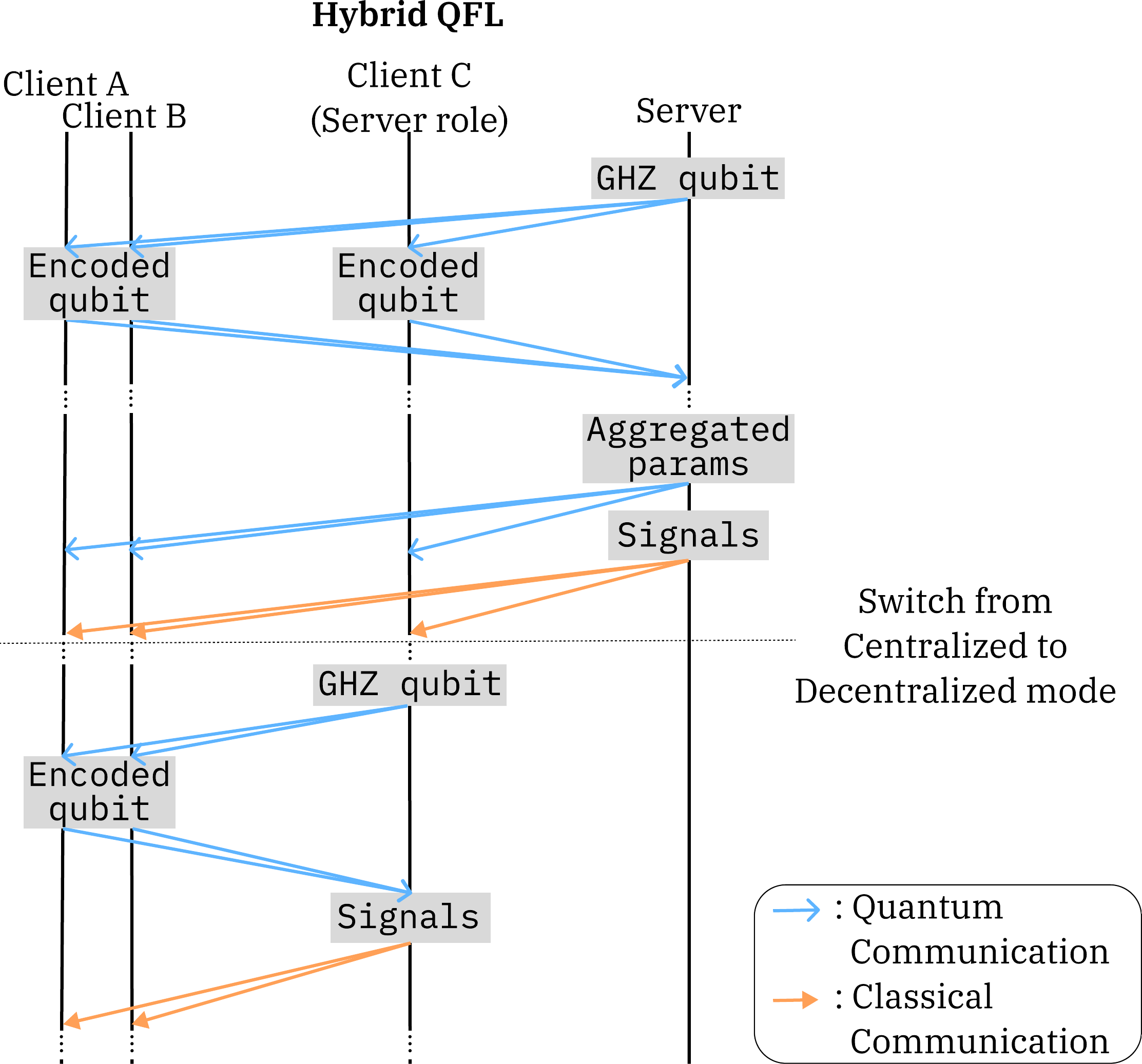}
    \caption{Hybrid QFL Protocol: The proposed Hybrid QFL integrates Centralized and Decentralized approaches sequentially. The transition is triggered based on validation performance.}
    \label{fig:hybridqfl}
\end{figure}

The overall flow of the proposed Hybrid QFL protocol is shown in Fig.~\ref{fig:hybridqfl}.

The proposed Hybrid QFL combines Centralized and Decentralized QFL in a sequential manner. In the initial phase, training follows the Centralized QFL procedure. After achieving a predefined validation performance threshold, the protocol transitions to the Decentralized QFL phase. The transition decision is determined by evaluating the global model accuracy at the server. Upon switching, subsequent training proceeds without redistributing the global model via quantum teleportation.

This design is motivated by two considerations. First, centralized redistribution via quantum teleportation introduces additional quantum transmissions. Per round, Decentralized QFL requires roughly $2\,N$ quantum transmissions for QSA (GHZ distribution and return), while Centralized QFL additionally requires about $N$ transmissions to redistribute the global model, resulting in roughly $3\,N$ transmissions. Transitioning to decentralized updates therefore reduces communication overhead. 
Second, while Decentralized QFL alone minimizes quantum transmissions, the absence of global model redistribution may slow convergence during early training stages. By initially leveraging centralized aggregation to accelerate convergence and subsequently switching to decentralized updates, the Hybrid protocol thereby balances communication efficiency and convergence performance.

\subsection{Model Architecture}

\begin{figure}[htbp]
    \centering
    \includegraphics[width=\linewidth]{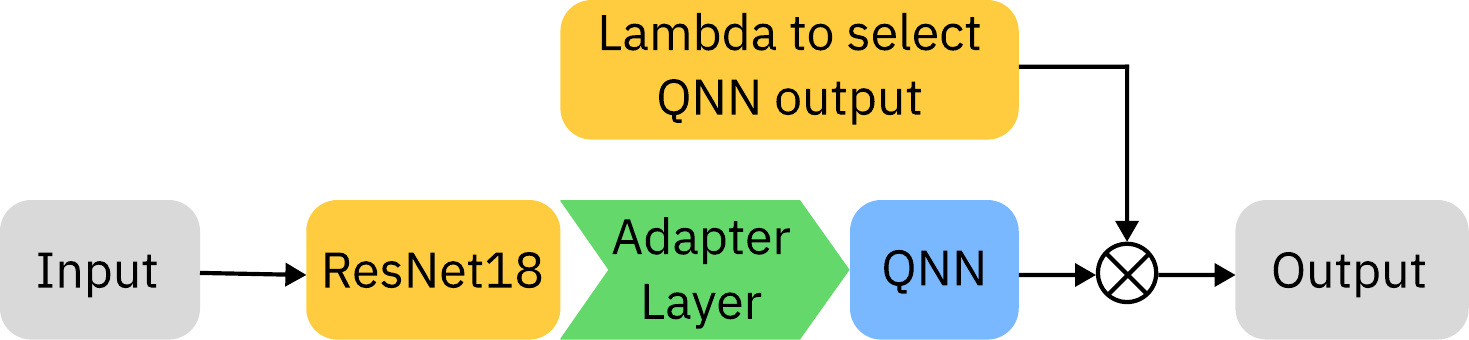}
    \caption{Overview of Model Architecture: A pre-trained ResNet18 is used as a feature extractor. The extracted features are passed through a classical linear Adapter layer and then into a 6-qubit Quantum Neural Network (QNN). The final classification output is computed as the weighted inner product between the QNN measurement expectations and a trainable vector $\lambda$.}
    \label{fig:modelarchitecture}
\end{figure}

We employ a hybrid classical-quantum model for binary medical-image classification (Fig.~\ref{fig:modelarchitecture}), distinguishing normal from abnormal inputs.

\begin{figure}[htbp]
    \centering
    \includegraphics[width=\linewidth]{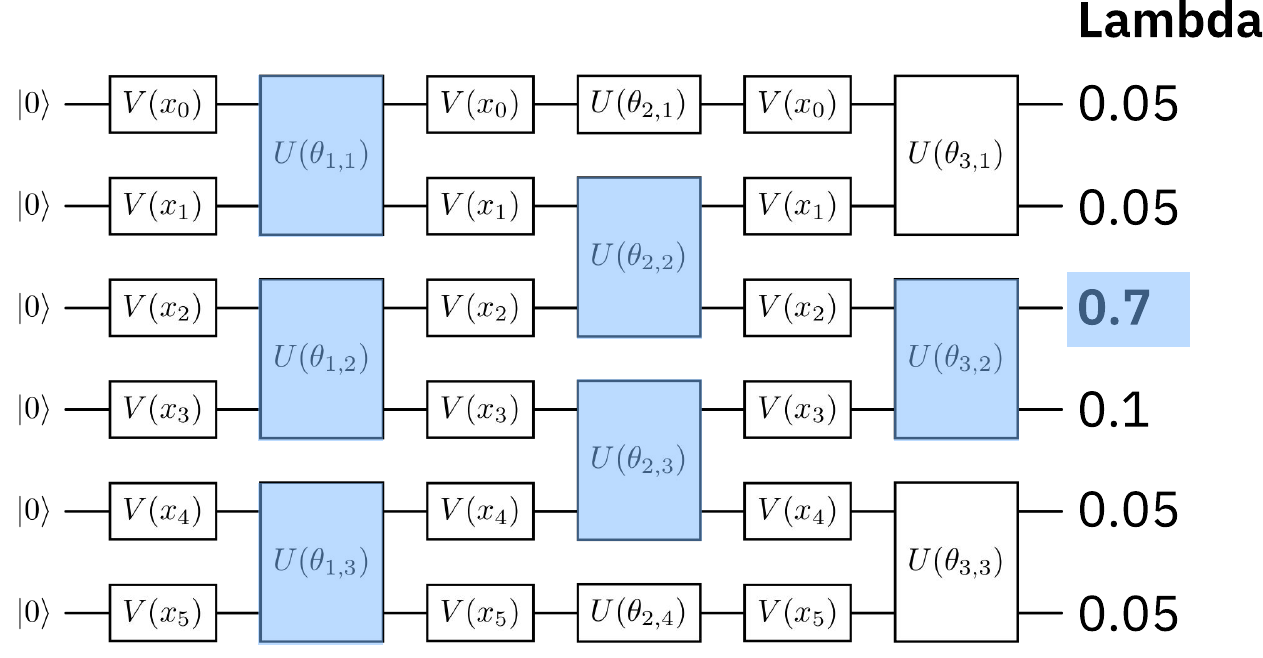}
    \caption{Light-cone feature selection: The dominant component of the trainable vector $\lambda$ determines the corresponding qubit output. Parameters contributing to this output form a cone-like structure in the quantum circuit, referred to as the light cone. Only parameters within this light cone are selected for aggregation.}
    \label{fig:lambdaselection}
\end{figure}

The input image is first processed using a transfer learning model based on ResNet18~\cite{he2016deep}, pretrained on ImageNet. The final fully connected layer of ResNet18 is removed and replaced with a single linear Adapter layer, which maps the extracted features to the input dimension required by the QNN. The transformed features are then fed into a 6-qubit QNN. As illustrated in Fig.~\ref{fig:lambdaselection}, the QNN adopts a data re-uploading strategy~\cite{P_rez_Salinas_2020} combined with a brick-like circuit structure~\cite{cerezo2021cost}. This layered entangling structure enables expressive nonlinear transformations while maintaining a limited number of qubits. The circuit topology is essential for the light-cone-based feature selection mechanism described below. The final scalar output is obtained by taking the inner product between the expected measurement values of each qubit and a trainable parameter vector $\lambda$, which is subsequently used for binary classification.

The trainable parameter vector $\lambda$ plays a dual role in the proposed architecture. After applying softmax normalization to $\lambda$, the final prediction is computed as the weighted sum of the expected measurement values of each qubit. Consequently, the magnitude of each component of $\lambda$ determines the relative contribution of the corresponding qubit to the classification result.

As shown in Fig.~\ref{fig:lambdaselection} this weighting mechanism induces a light-cone-based feature selection process.

Because each output qubit depends only on a restricted subset of circuit parameters in the brick-like topology, selecting the dominant component of $\lambda$ effectively determines the associated light cone in the circuit. Only parameters within this light cone are aggregated and transmitted, while the remaining parameters are excluded from communication. This mechanism reduces the number of transmitted parameters without altering the overall circuit structure.

Light-cone feature selection is applied dynamically at the beginning of each global round. Specifically, the server determines the set of parameters to be aggregated based on the $\lambda$ vector of the current global model. After applying softmax normalization to $\lambda$, the qubit with the highest weight is identified, and all circuit parameters within its causal light cone are selected for aggregation. The adapter layer parameters and $\lambda$ itself are always included regardless of the light-cone selection.

In the first round, $\lambda$ is initialized uniformly as $1/n_{\text{qubits}}$, resulting in equal softmax weights across all qubits. From the second round onward, $\lambda$ is updated via FedAvg aggregation of the raw (pre-softmax) parameter values across clients. As training progresses and the model specializes, $\lambda$ may develop a dominant component, causing the light cone to concentrate on a specific qubit. Both the server and each client apply the same selection criterion using the shared global model state, ensuring that the aggregated parameter indices are consistent across all participants in each round.

The main architectural and training hyperparameters used in this study are summarized in Table~\ref{table:hyperparameters}.

\begin{table}[htbp]
\centering
\caption{Hyperparameters and Configuration}
\label{table:hyperparameters}
\begin{tabularx}{0.5\textwidth}{@{} l l X @{}}
\toprule
\textbf{Category} & \textbf{Hyperparameter / Setting} & \textbf{Value} \\
\midrule
\multirow{4}{*}{Classical Network} 
 & Base Model & ResNet18 (ImageNet) \\
 & Input Resolution & $224 \times 224 \times 3$ \\
 & Intermediate Feature Dimension & 512 \\
\midrule
\multirow{6}{*}{Quantum Circuit} 
 & Number of Qubits & 6 \\
 & Number of Layers & 3 \\
 & Entanglement Pattern & Brickwork \\
 & Measurement & Pauli-Z Expectation \\
 & Backend Device & \texttt{lightning.qubit} (PennyLane) \\
 & Differentiation Method & Adjoint \\
\midrule
\multirow{4}{*}{Training} 
 & Optimizer & Adam \\
 & Learning Rate & $0.001$ \\
 & Batch Size & 16 \\
 & Loss Function & BCE Loss \\
\bottomrule
\end{tabularx}
\end{table}

Unless otherwise specified, these hyperparameters were fixed throughout the following experiments (Section~\ref{sec:resultsanddiscussion}) to isolate the effects of the proposed communication strategy and light-cone-based feature selection.

\section{Experimental Setup}
\label{sec:experimentalsetup}
\subsection{Datasets}

\begin{figure}[htbp]
    \centering
    \includegraphics[width=\linewidth]{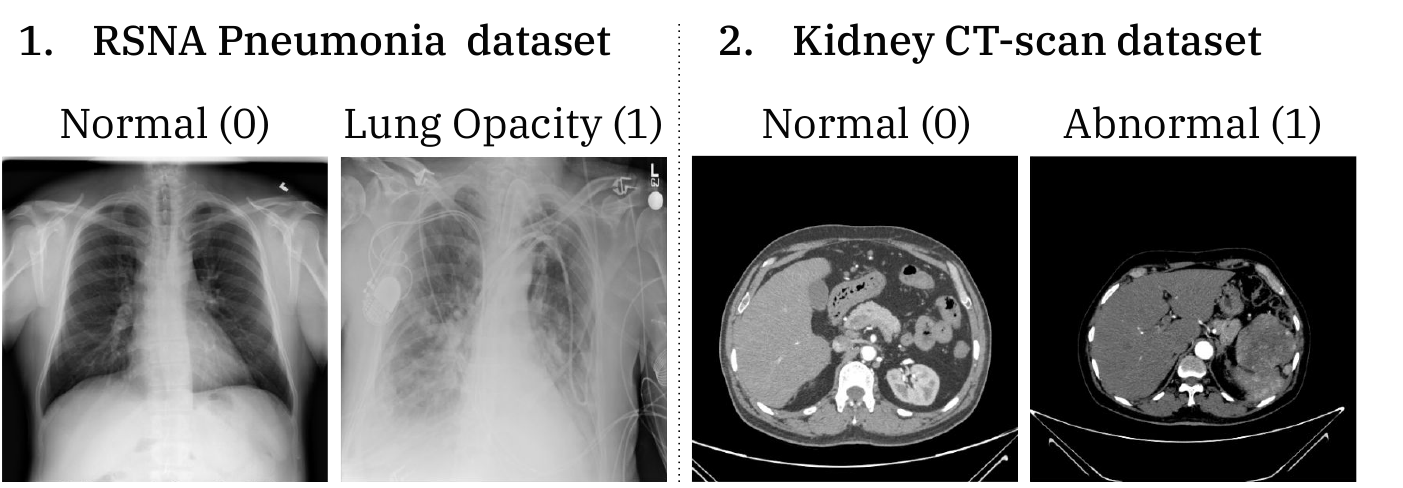}
    \caption{Datasets: Representative samples from the RSNA Chest X-ray dataset~\cite{shih2019augmenting} and the Kidney CT-scan dataset~\cite{islam2022vision}.}
    \label{fig:dataset}
\end{figure}

To validate the proposed framework in the context of medical diagnosis, we use two publicly available datasets: the RSNA Chest X-ray dataset~\cite{shih2019augmenting} and the Kidney CT-scan dataset~\cite{islam2022vision}, both treated as binary classification tasks (Fig.~\ref{fig:dataset}).

\begin{figure}[htbp]
    \centering
    \includegraphics[width=\linewidth]{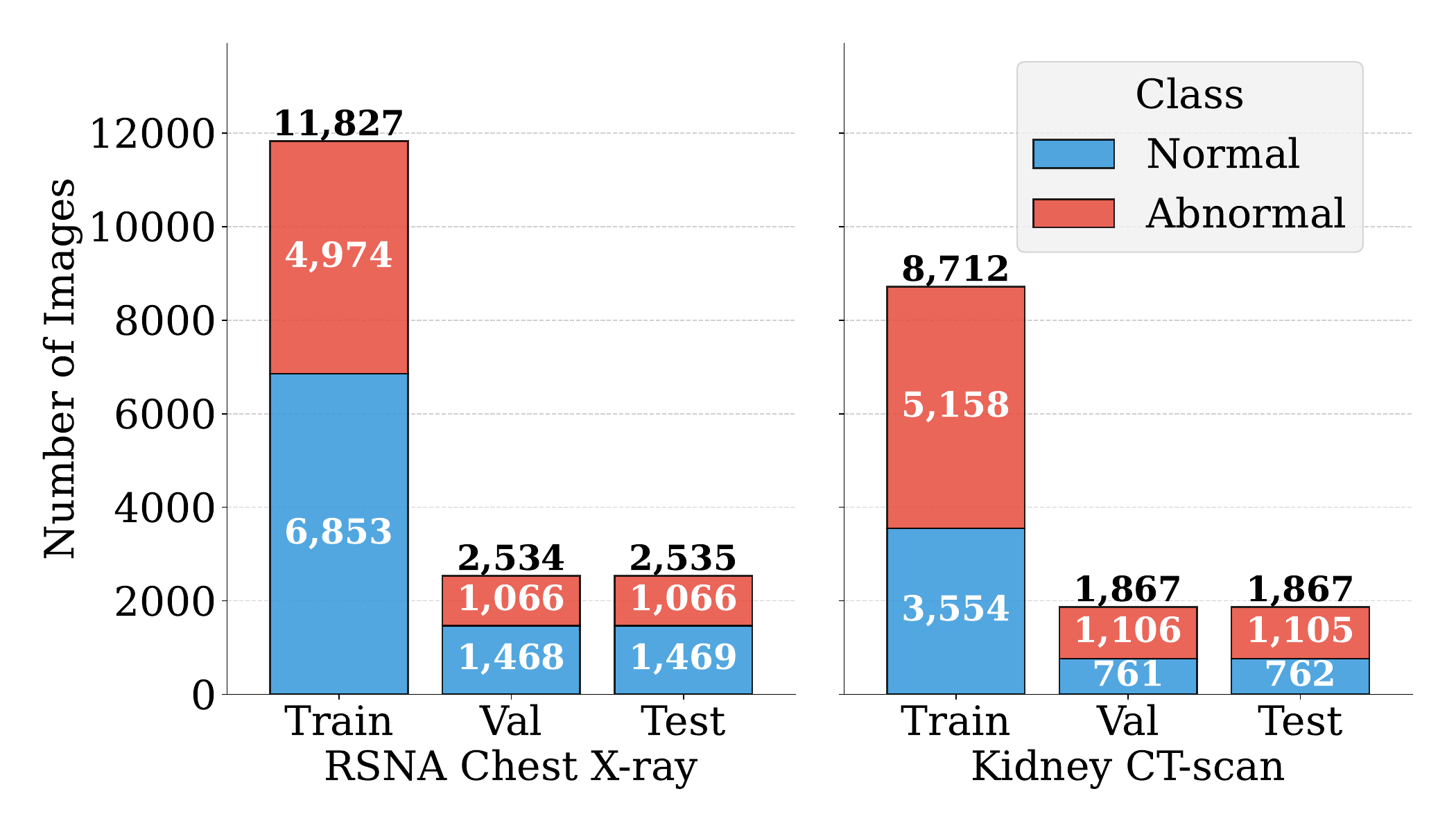}
    \caption{Distribution of datasets across Train, Validation, and Test splits. Training data are distributed among clients, while validation and test data are used to evaluate global model performance.}
    \label{fig:distributionofdataset}
\end{figure}

The RSNA dataset consists of chest X-ray images labeled as normal or abnormal, while the Kidney CT-scan dataset contains CT images classified according to the presence or absence of renal disease. Fig.~\ref{fig:distributionofdataset} illustrates the distribution of the data into training, validation, and test splits.

The number of federated clients is set to 3, 5, and 7 to evaluate scalability. The detailed client-wise data distributions for the RSNA and Kidney CT-scan datasets are provided in Appendix~\ref{sec:appendix}.

The non-IID setting in this study reflects quantity skew across clients, where class label ratios remain approximately uniform but the number of samples per client varies significantly (e.g., 232 to 5,465 samples in the 7-client RSNA setting). The degree of heterogeneity can be characterized by the sample size imbalance ratio of approximately $16:1$ between the largest and smallest clients.

\subsection{Depolarizing Noise Model}

To evaluate robustness against quantum channel noise, we adopt a depolarizing noise model. The depolarizing channel transforms a density matrix $\rho$ in a $d$-dimensional system as $\mathcal{E}(\rho) = (1 - p)\rho + p \frac{I}{d}$, where $p$ denotes the depolarizing rate and $I/d$ represents the maximally mixed state.

For a pure state $\rho = \ket{\psi}\bra{\psi}$, the fidelity between the noisy and ideal states becomes $F = 1 - \left(1 - \frac{1}{d}\right)p$. This formulation allows us to quantitatively control the noise strength and analyze its impact on model performance.

\subsection{Implementation}

The classical layers of the machine learning model were constructed using PyTorch~\cite{paszke2017automatic}, and the quantum layers were implemented using PennyLane~\cite{bergholm2022pennylaneautomaticdifferentiationhybrid}. Quantum Secure Aggregation, quantum teleportation, and classical communication were simulated using the quantum network simulator NetSquid~\cite{Coopmans_2021}.

\subsection{Experiment Details}

To evaluate the proposed framework, we conduct three experiments targeting (i) communication reduction, (ii) architectural efficiency, and (iii) robustness against quantum channel noise. Unless otherwise specified, all experiments share the common settings summarized in Table~\ref{table:common_settings}.

\begin{table}[htbp]
    \centering
    \caption{Common Experimental Settings}
    \label{table:common_settings}
    \begin{tabularx}{0.48\textwidth}{@{} l X @{}}
        \toprule
        \textbf{Setting} & \textbf{Description} \\
        \midrule
        Datasets & RSNA chest X-ray, Kidney CT-scan \\
        Training Schedule & 30 global training rounds (1 local epoch per round) \\
        Data Distribution & Non-IID \\
        \bottomrule
    \end{tabularx}
\end{table}

\textbf{Experiment 1:} Effectiveness of Feature Selection: 
This experiment evaluates whether light-cone-based feature selection can reduce quantum transmissions while maintaining model accuracy.

We compare the following aggregation strategies across 3, 5, and 7 clients:

\begin{enumerate}
    \item Full aggregation: All trainable parameters are aggregated.
    \item Random partial aggregation: Adapter and $\lambda$ parameters are fully aggregated, while QNN parameters are randomly subsampled.
    \item Light-cone aggregation: Adapter and $\lambda$ parameters are aggregated, together with only the QNN parameter group corresponding to the dominant $\lambda$ component.
\end{enumerate}

\textbf{Experiment 2:} Effectiveness of Hybrid QFL: 
This experiment evaluates whether the Hybrid QFL framework improves the trade-off between communication cost and convergence stability.

The transition from Centralized to Decentralized QFL is triggered when the global validation accuracy exceeds a threshold $\tau$, determined empirically from preliminary training curves.

We compare the following architectures across 3, 5, and 7 clients:

\begin{enumerate}
    \item Centralized QFL
    \item Decentralized QFL
    \item Hybrid QFL (Switching threshold $\tau = 0.85$ for RSNA and $0.9$ for Kidney)
\end{enumerate}

\textbf{Experiment 3:} Impact of Quantum Communication Noise and Error Correction: 
This experiment assesses robustness under realistic quantum channel noise. Depolarizing noise with $p \in \{0.0001, 0.00025, 0.0005, 0.00075, 0.001\}$ is applied to all quantum channels. Using a 3-client configuration, we evaluate:

\begin{enumerate}
    \item Centralized QFL (Full and Light-cone aggregation)
    \item Decentralized QFL
\end{enumerate}

To examine error mitigation capability, we further apply the Steane code at depolarizing noise with $p \in \{0.001, 0.0025, 0.005, 0.0075, 0.01\}$. Qubits are encoded prior to transmission, error correction is performed upon reception, and decoded states are used for aggregation.

\section{Results and Discussion}
\label{sec:resultsanddiscussion}

\begin{figure*}[htbp]
\centering
\begin{subfigure}[t]{0.49\textwidth}
    \centering
    \includegraphics[width=1\linewidth]{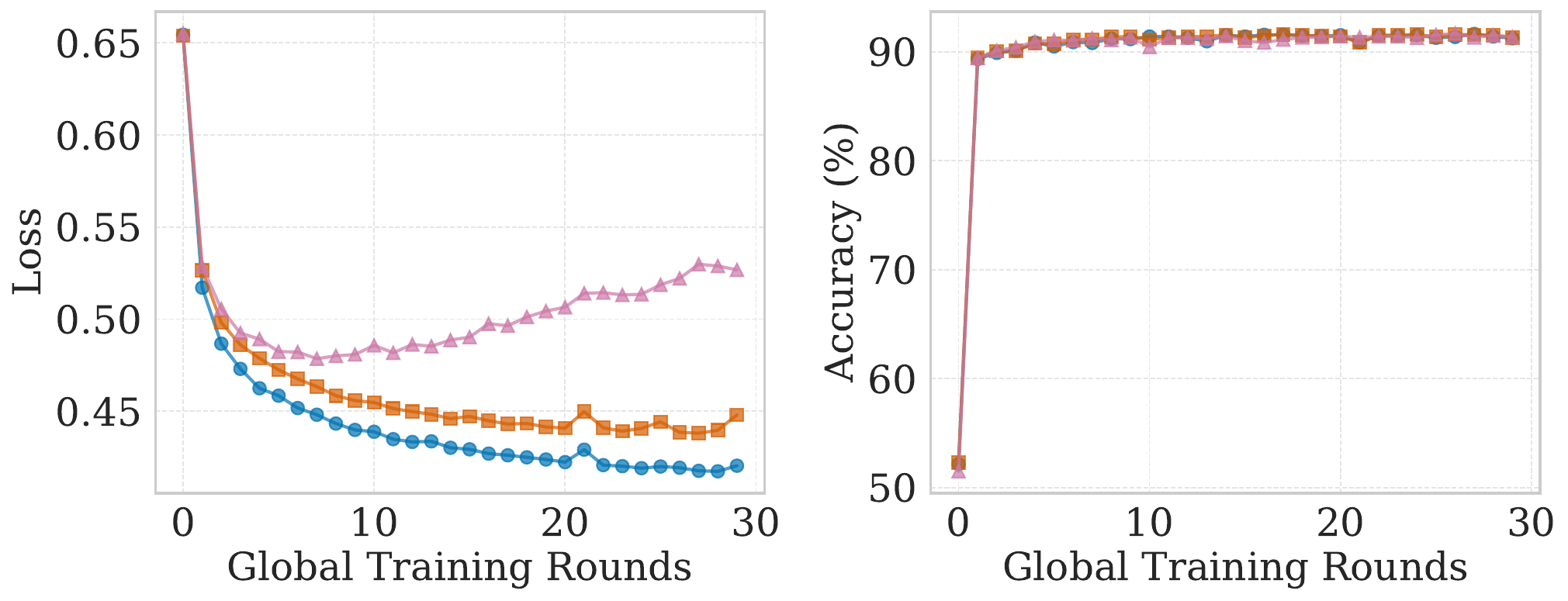}
    \caption{}
    \label{fig:exp1_rsna}
\end{subfigure}
\hfill
\begin{subfigure}[t]{0.49\textwidth}
    \centering
    \includegraphics[width=1\linewidth]{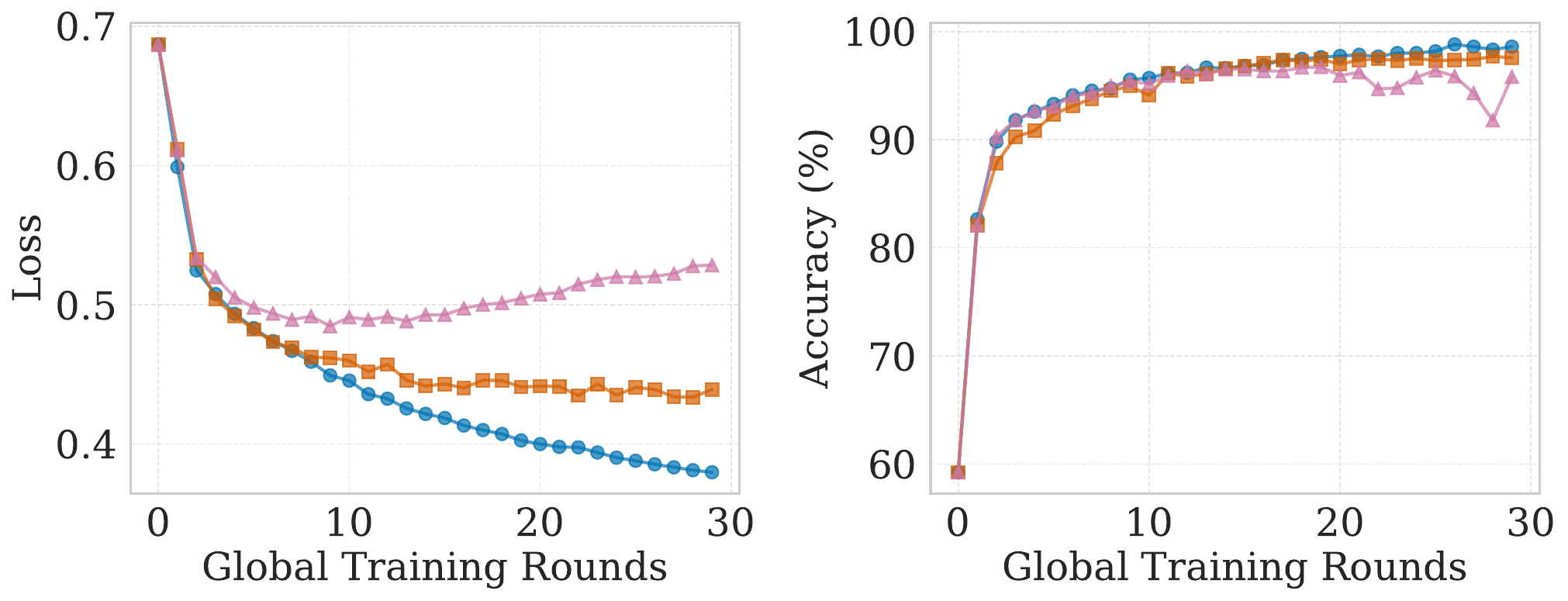}
    \caption{}
    \label{fig:exp1_ctkidney}
\end{subfigure}
 
\vspace{0.5em}
 
\begin{subfigure}[t]{0.49\textwidth}
    \centering
    \includegraphics[width=1\linewidth]{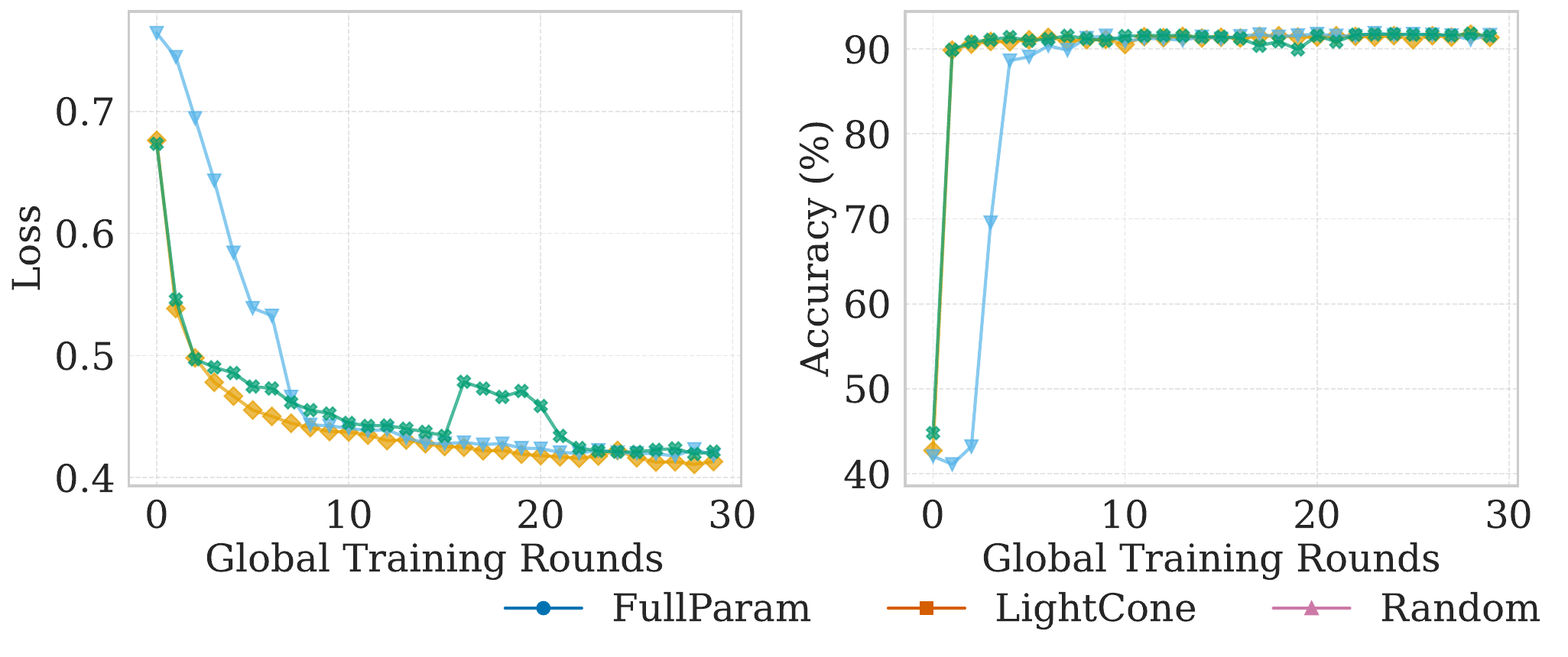}
    \caption{}
    \label{fig:exp2_rsna}
\end{subfigure}
\hfill
\begin{subfigure}[t]{0.49\textwidth}
    \centering
    \includegraphics[width=1\linewidth]{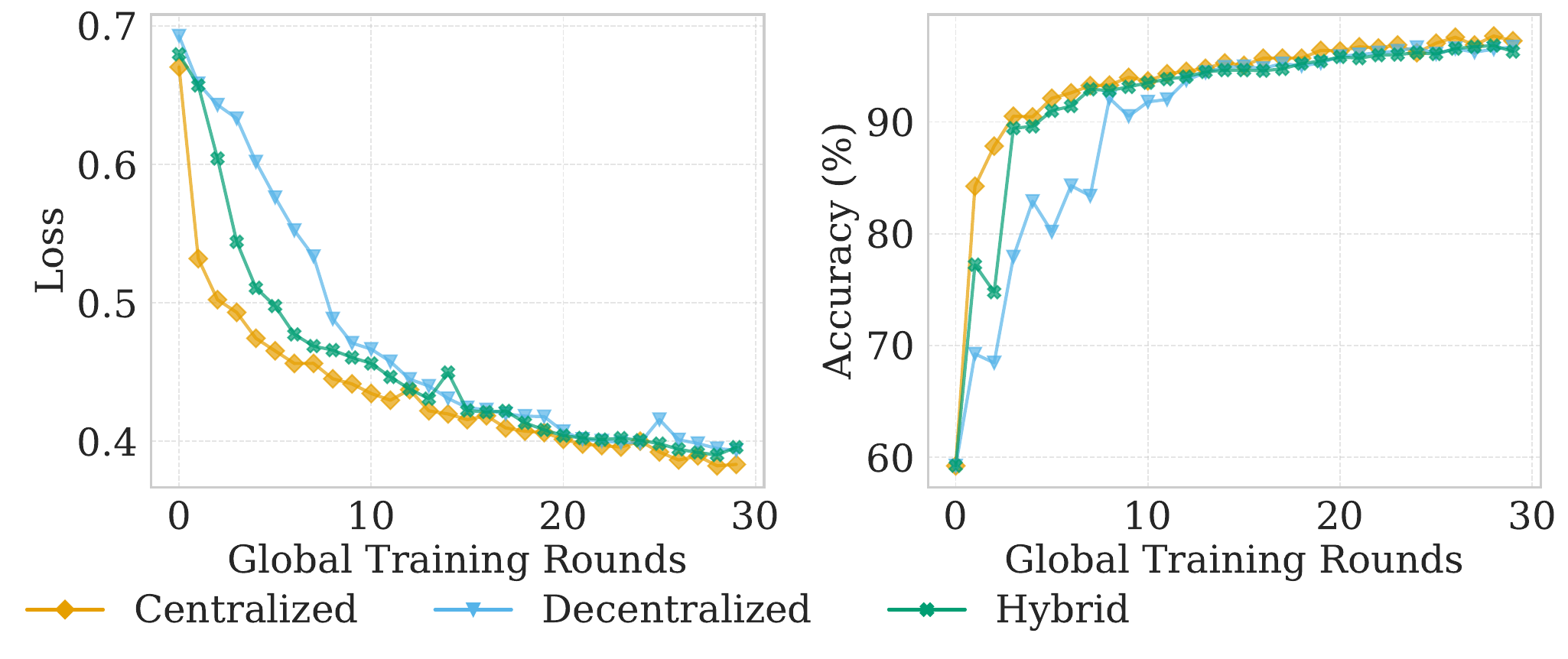}
    \caption{}
    \label{fig:exp2_ctkidney}
\end{subfigure}
\caption{Comparison of aggregation methods (FullParam, Light-cone, Random) and network structures (Centralized, Decentralized, Hybrid). The horizontal axis shows Global Training Rounds (each representing a single federated learning cycle). For each subfigure, the left and right vertical axes denote Loss and Accuracy, respectively. (a) Aggregation methods, 7 Clients, RSNA Chest X-ray dataset. (b) Aggregation methods, 7 Clients, Kidney CT-scan dataset. (c) Network structures, 7 Clients, RSNA Chest X-ray dataset. (d) Network structures, 7 Clients, Kidney CT-scan dataset.}
\label{fig:exp1andexp2}
\end{figure*}

\subsection*{Experiment 1: Effectiveness of Light-cone Feature Selection}

\begin{table}[htbp]
\centering
\caption{Test accuracy (\%) of aggregation methods.}
\label{table:aggregation}
\begin{tabularx}{0.46\textwidth}{@{} lcccc @{}}
\toprule
\textbf{Dataset} & \textbf{Clients} & \textbf{FullParam} & \textbf{LightCone} & \textbf{Random} \\
\midrule
\multirow{3}{*}{RSNA Chest X-ray} & 3 & 91.1 & 91.2 & 91.0 \\
                                   & 5 & 90.7 & 90.6 & 90.7 \\
                                   & 7 & 91.1 & 90.9 & 90.8 \\
\midrule
\multirow{3}{*}{Kidney CT-scan}   & 3 & 97.1 & 96.4 & 97.3 \\
                                   & 5 & 98.7 & 97.3 & 98.2 \\
                                   & 7 & 97.9 & 97.1 & 95.9 \\
\bottomrule
\end{tabularx}
\end{table}

Full aggregation achieves competitive accuracy and stable convergence across both datasets and client counts, serving as an upper bound for performance. As shown in Table~\ref{table:aggregation}, Light-cone aggregation closely matches full aggregation on the RSNA Chest X-ray dataset and remains competitive on the Kidney CT-scan dataset even as the client count increases, while reducing the number of aggregated parameters. In contrast, random partial aggregation yields greater instability, especially with larger client counts (e.g., 95.9\% at 7 clients on Kidney CT-scan), underscoring the value of structured parameter selection. This behavior is also visible in Fig.~\ref{fig:exp1_rsna} and~\ref{fig:exp1_ctkidney}, where the loss curves for the 7-client setting confirm that Light-cone aggregation tracks Full aggregation closely throughout training, while Random aggregation exhibits more fluctuation.

To quantify the communication impact, we evaluate the total quantum communication volume during training. The model contains 3120 trainable parameters, including 36 QNN parameters, and training is conducted over 30 global rounds. The total communication volume per training run under Centralized QFL is defined as $R =  3\,TNMP$, consisting of QSA aggregation ($2\,NMP$ per round) and global model redistribution via quantum teleportation ($NMP$ per round); the derivation of these per-round costs is given in Experiment 2. $T = 30$ denotes the number of global rounds, $N$ the number of clients, $M$ the number of measurement shots per parameter, and $P$ denotes the number of aggregated parameters per round. Table~\ref{table:communication_exp1} summarizes the communication volume for each aggregation strategy.

\begin{table}[htbp]
\centering
\caption{Quantum communication volume per training run for each aggregation strategy ($T = 30$). $R$ is computed as $3TNMP$, where $P$ is shown in the preceding column.}
\label{table:communication_exp1}
\begin{tabularx}{0.28\textwidth}{@{} lrr @{}}
\toprule
\textbf{Strategy} & \textbf{$P$} & \textbf{$R$} \\
\midrule
Full aggregation   & 3120 & $280{,}800\,NM$ \\
Light-cone         & 3104 & $279{,}360\,NM$ \\ \midrule
Saving             & 16   & $1{,}440\,NM$ \\
\bottomrule
\end{tabularx}
\end{table}

Although the reduction is modest at the current circuit scale (approximately 0.5\%), the absolute saving grows as the circuit scales. In the current 6-qubit, 3-layer brickwork configuration, the light-cone selects 20 of 36 QNN parameters for aggregation. If the QNN is scaled to 12 qubits and 6 layers while preserving the brickwork topology and a single dominant $\lambda$ component, the QNN parameter count grows as $Q \times L$ (from 36 to approximately 144), and the excluded parameter count grows proportionally to approximately 64, yielding a saving of approximately $5{,}760\,NM$ quantum transmissions per training run, a fourfold increase over the current setting. This suggests that light-cone-based parameter reduction becomes increasingly effective as QNN scale grows.

\subsection*{Experiment 2: Effectiveness of Hybrid QFL}

\begin{table}[htbp]
\centering
\caption{Test accuracy (\%) of network structures.}
\label{table:network}
\begin{tabularx}{0.48\textwidth}{@{} lcccc @{}}
\toprule
\textbf{Dataset} & \textbf{Clients} & \textbf{Centralized} & \textbf{Decentralized} & \textbf{Hybrid} \\
\midrule
\multirow{3}{*}{RSNA Chest X-ray} & 3 & 90.5 & 90.9 & 90.8 \\
                                   & 5 & 90.9 & 91.1 & 90.7 \\
                                   & 7 & 90.9 & 90.5 & 91.0 \\
\midrule
\multirow{3}{*}{Kidney CT-scan}   & 3 & 98.6 & 98.8 & 97.9 \\
                                   & 5 & 97.7 & 96.8 & 97.6 \\
                                   & 7 & 97.5 & 97.2 & 97.5 \\
\bottomrule
\end{tabularx}
\end{table}

Table~\ref{table:network} presents the test accuracy of Centralized, Decentralized, and Hybrid QFL. Across both datasets, the three structures achieve comparable final accuracy, though their convergence behavior differs. As can be seen in Fig.~\ref{fig:exp2_rsna} and~\ref{fig:exp2_ctkidney}, Centralized QFL converges fastest, Hybrid QFL follows closely, and Decentralized QFL is slowest. Hybrid QFL retains most of the convergence benefits of centralization while reducing communication overhead, and is more stable than Decentralized QFL.

The Kidney CT-scan dataset further illustrates this effect: increasing the number of clients amplifies instability in Decentralized QFL (97.2\% at 7 clients), whereas Hybrid QFL maintains accuracy on par with Centralized QFL (97.5\%). The training curves in Fig.~\ref{fig:exp2_ctkidney} corroborate this, showing that Decentralized QFL exhibits more pronounced loss fluctuations in the Kidney setting compared to the RSNA dataset. Because Decentralized QFL updates models without global redistribution, the aggregated model becomes more susceptible to client-specific data bias under stronger heterogeneity. By performing centralized aggregation in the early training stage and switching to decentralized mode afterward, Hybrid QFL mitigates this instability while reducing total communication cost.

To quantify the communication trade-off, Table~\ref{table:communication_exp2} summarizes the per-round and total quantum transmission cost for each architecture.

\begin{table}[htbp]
\centering
\caption{Quantum communication cost per round and total over $T = 30$ rounds.}
\label{table:communication_exp2}
\begin{tabularx}{0.48\textwidth}{@{} lrr @{}}
\toprule
\textbf{Architecture} & \textbf{Per-round cost} & \textbf{Total ($T=30$)} \\
\midrule
Centralized    & $3\,NMP$ & $90\,NMP$ \\
Decentralized  & $2\,NMP$ & $60\,NMP$ \\
Hybrid & $3\,NMP$ / $2\,NMP$ (switch) & $\{3t + 2(T-t)\}\,NMP$ \\
\bottomrule
\end{tabularx}
\end{table}

In each round, both the QSA-based aggregation and the teleportation-based redistribution involve repeated quantum transmissions to account for shot-noise estimation.

In QSA, each GHZ state is consumed by a single measurement, so the server must repeat the protocol $M$ times per parameter to obtain a statistically reliable estimate of the aggregated value, yielding $R_a = 2\,NMP$ transmissions per round (GHZ distribution and return for $N$ clients, $P$ parameters, and $M$ shots).

Similarly, in redistribution, the server prepares an encoded quantum state of each aggregated parameter and teleports $M$ copies to each client so that the client can reconstruct the global parameter to comparable shot-noise precision via measurement, yielding $R_b = NMP$ transmissions per round.

Consequently, the per-round costs are $R_{\text{Centralized}} = R_a + R_b = 3\,NMP$ and $R_{\text{Decentralized}} = R_a = 2\,NMP$. For Hybrid QFL transitioning from centralized to decentralized mode at round $t$, the total cost is $R_{\text{Hybrid}} = \{3t + 2(T - t)\}\,NMP$, yielding a reduction of $(T - t)\,NMP$ compared to Centralized QFL. With $t = 2$ or $3$, this corresponds to a saving of approximately $27\,NMP$ to $28\,NMP$ quantum transmissions, demonstrating that Hybrid QFL substantially reduces communication cost while retaining most convergence advantages.

The switching threshold $\tau$ was determined empirically by inspecting preliminary training curves. Specifically, $\tau = 0.85$ for RSNA and $\tau = 0.9$ for Kidney CT-scan were selected at the point where the validation accuracy curve visually plateaued under Centralized QFL (as observable in Fig.~\ref{fig:exp2_rsna} and ~\ref{fig:exp2_ctkidney}), indicating that the global model had reached a stable initialization sufficient for decentralized updates. A systematic sensitivity analysis of $\tau$ is left as future work.

\subsection*{Experiment 3: Impact of Depolarizing Noise and Quantum Error Correction}

\begin{figure}[htbp]
\centering

\begin{subfigure}{0.24\textwidth}
    \centering
    \includegraphics[width=\linewidth]{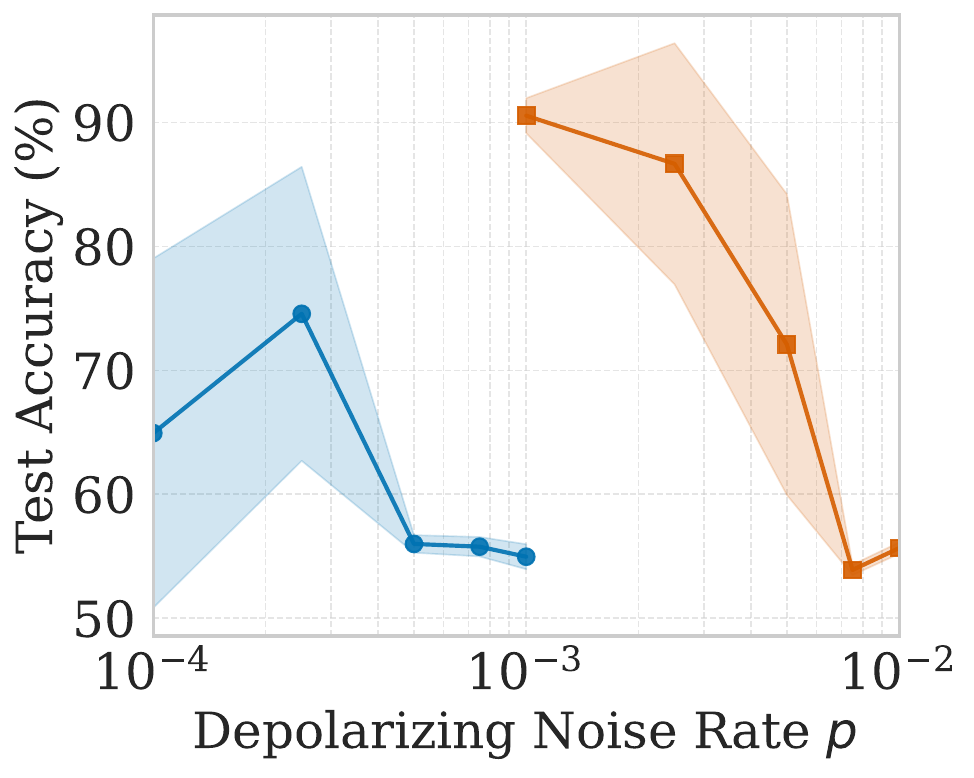}
    \caption{}
    \label{fig:exp3_centralized_rsna}
\end{subfigure}
\hfill
\begin{subfigure}{0.24\textwidth}
    \centering
    \includegraphics[width=\linewidth]{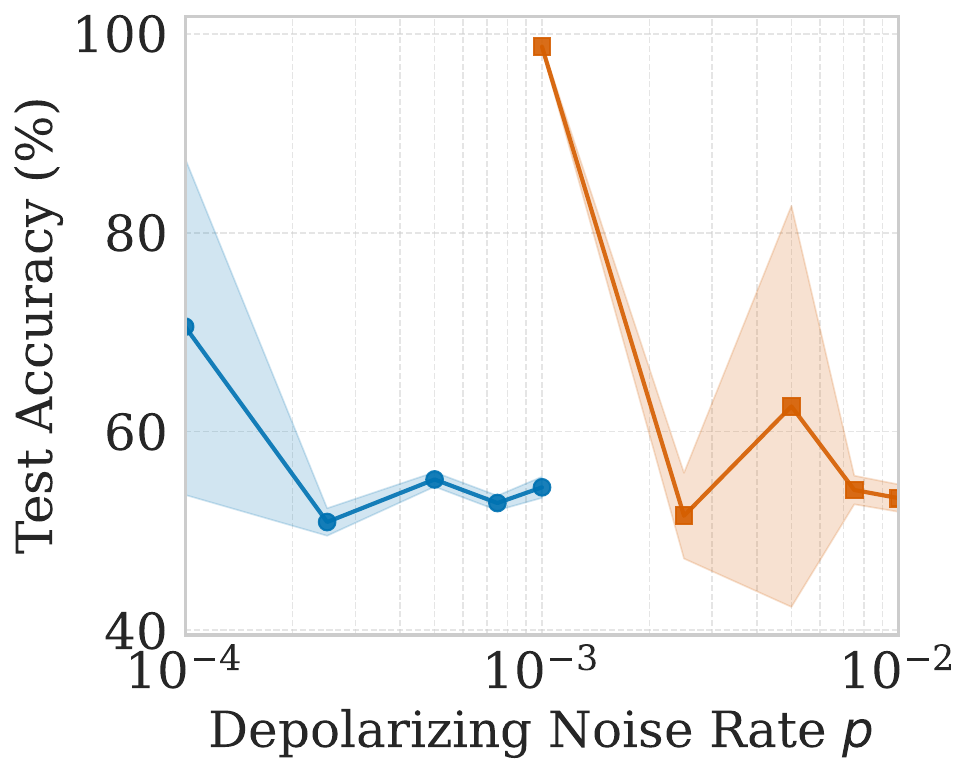}
    \caption{}
    \label{fig:exp3_centralized_kidney}
\end{subfigure}

\vspace{0.5em}

\begin{subfigure}{0.24\textwidth}
    \centering
    \includegraphics[width=\linewidth]{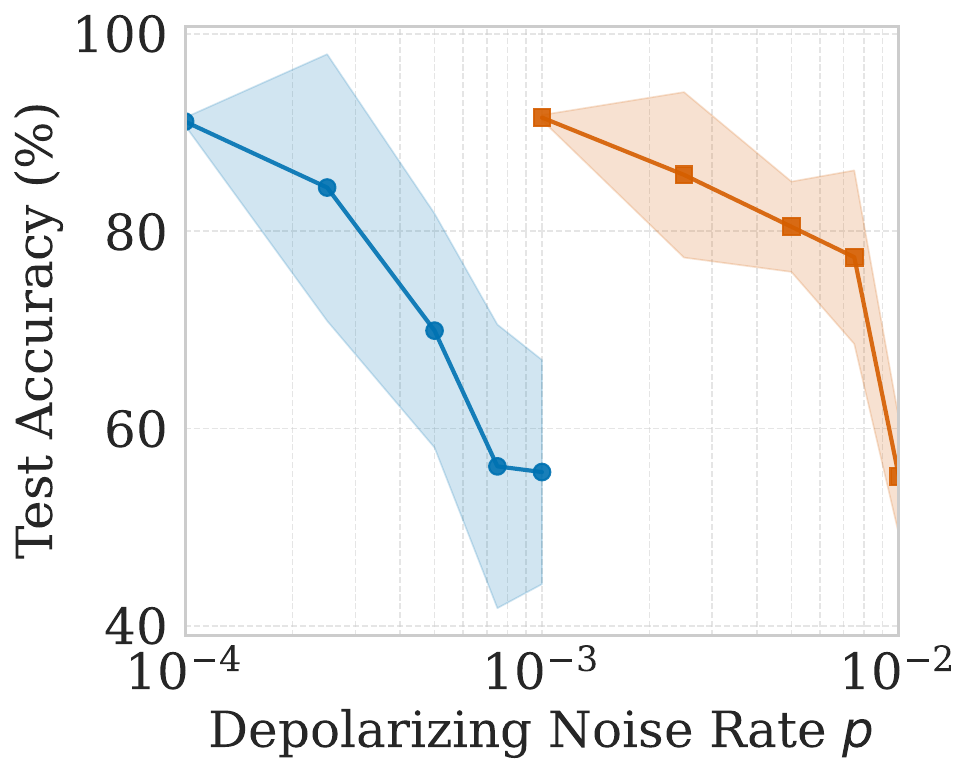}
    \caption{}
    \label{fig:exp3_decentralized_rsna}
\end{subfigure}
\hfill
\begin{subfigure}{0.24\textwidth}
    \centering
    \includegraphics[width=\linewidth]{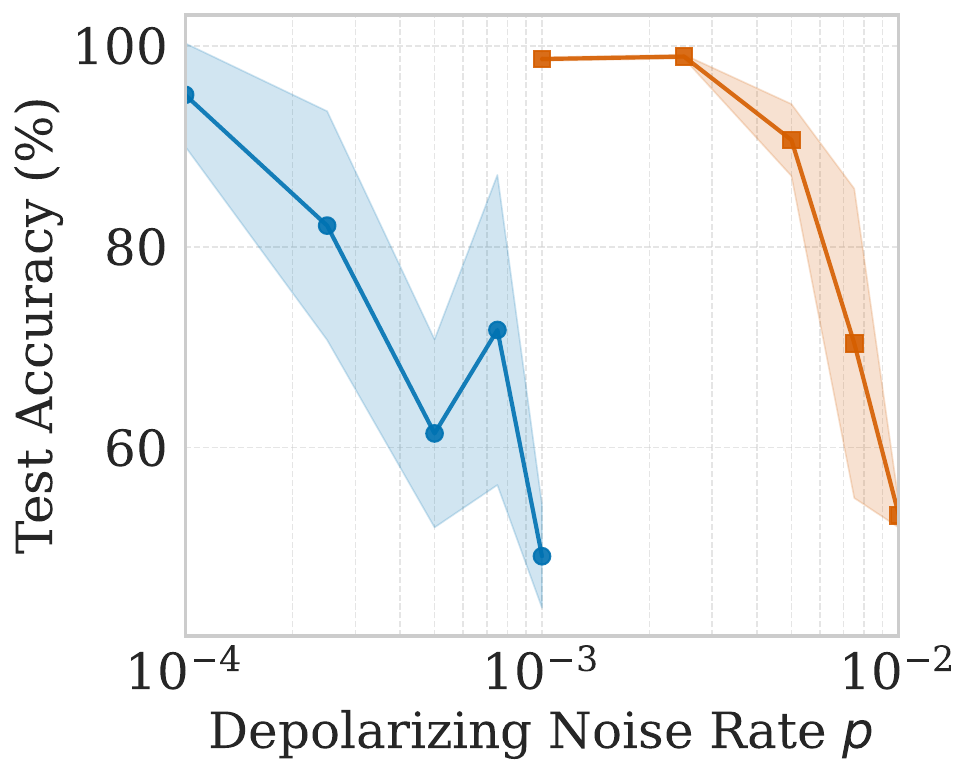}
    \caption{}
    \label{fig:exp3_decentralized_kidney}
\end{subfigure}

\vspace{0.5em}

\begin{subfigure}{0.24\textwidth}
    \centering
    \includegraphics[width=\linewidth]{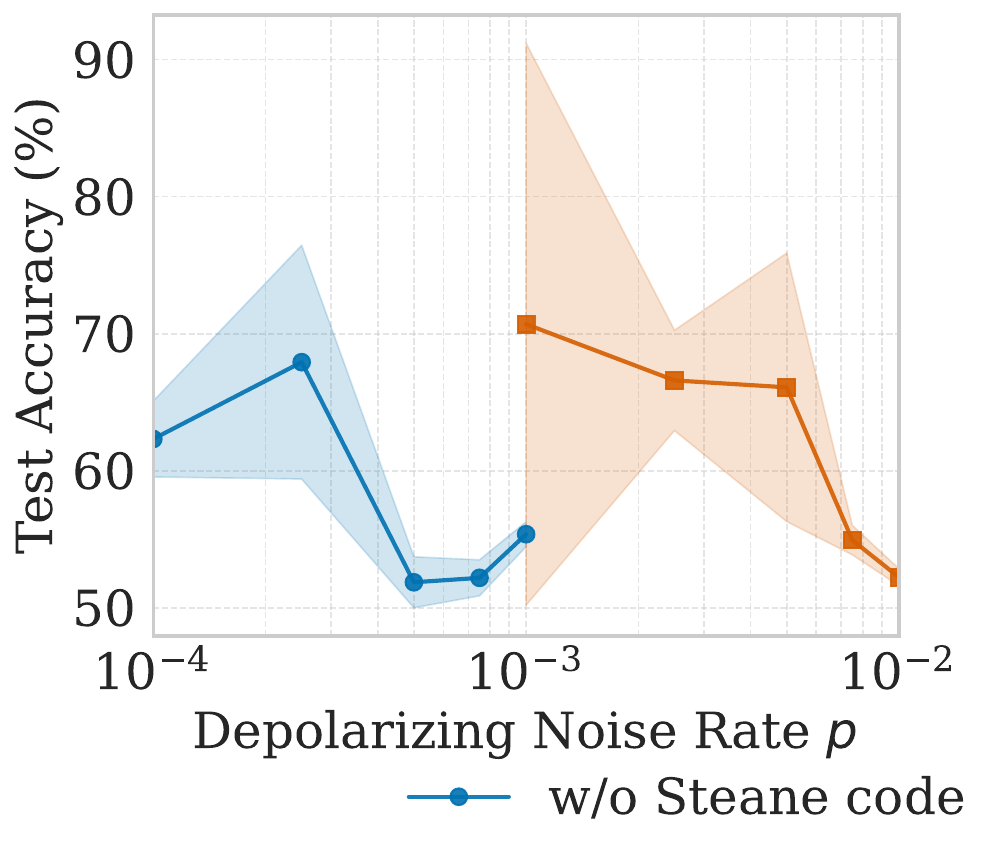}
    \caption{}
    \label{fig:exp3_lightcone_rsna}
\end{subfigure}
\hfill
\begin{subfigure}{0.24\textwidth}
    \centering
    \includegraphics[width=\linewidth]{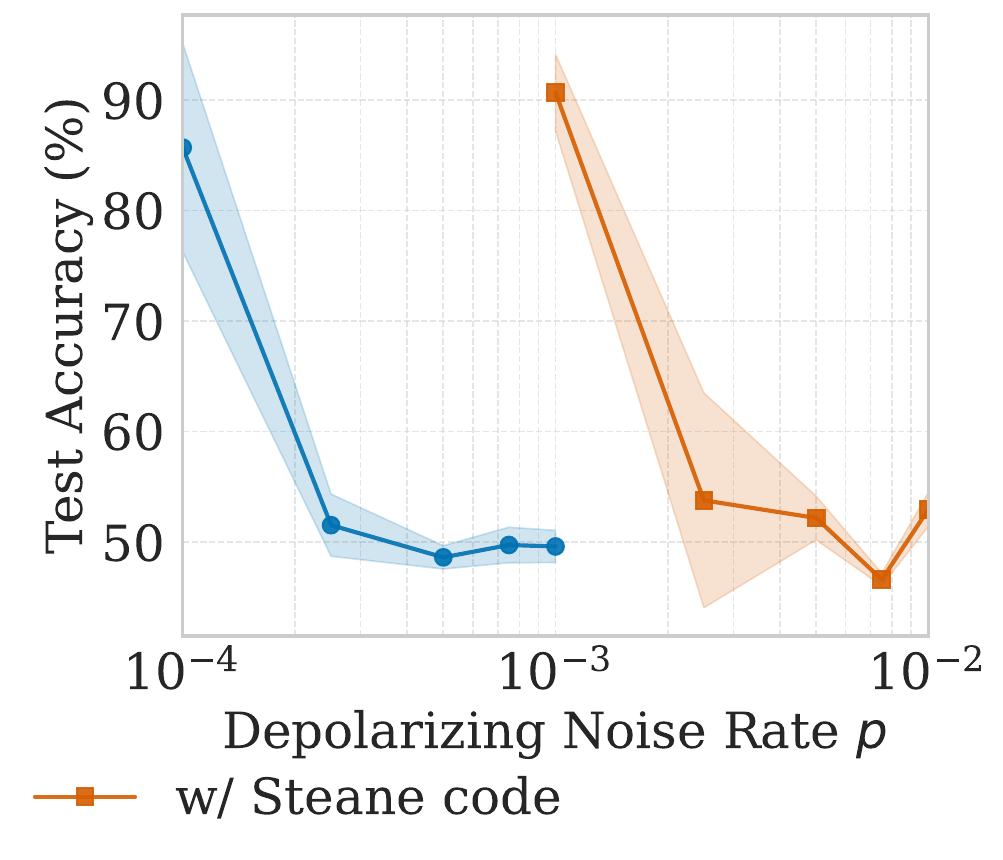}
    \caption{}
    \label{fig:exp3_lightcone_kidney}
\end{subfigure}

\caption{Impact of depolarizing noise on test accuracy. Results are shown for noise rates $p \in [10^{-4}, 10^{-3}]$ without Steane encoding (blue curves), and for $p \in [10^{-3}, 10^{-2}]$ with Steane encoding (orange curves). The horizontal axis shows the depolarizing noise rate $p$ on a logarithmic scale, and the vertical axis shows test accuracy (\%). Shaded regions indicate the standard deviation computed over the last 10 global training rounds.
(a) Centralized QFL, RSNA Chest X-ray dataset. 
(b) Centralized QFL, Kidney CT-scan dataset. 
(c) Decentralized QFL, RSNA Chest X-ray dataset. 
(d) Decentralized QFL, Kidney CT-scan dataset. 
(e) Centralized QFL with Light-cone aggregation, RSNA Chest X-ray dataset. 
(f) Centralized QFL with Light-cone aggregation, Kidney CT-scan dataset.}
\label{fig:exp3}
\end{figure}

Fig.~\ref{fig:exp3} shows test accuracy as a function of depolarizing noise rate $p$ for Centralized QFL, Decentralized QFL, and Centralized QFL with Light-cone aggregation, with and without Steane code error correction.

\subsubsection*{Noise Sensitivity}

Across all configurations, learning is severely degraded at $p = 10^{-3}$: accuracy stays near chance level and exhibits high variance. At $p = 10^{-4}$, learning becomes partially viable but exhibits oscillatory behavior, indicating that noise accumulation intermittently disrupts optimization.

Decentralized QFL (Fig.~\ref{fig:exp3_decentralized_rsna} and~\ref{fig:exp3_decentralized_kidney}) demonstrates comparatively greater robustness under noisy conditions. Even at $p = 10^{-3}$, partial learning progress can be observed in both datasets, particularly for the Kidney dataset with Steane encoding (Fig.~\ref{fig:exp3_decentralized_kidney}), where accuracy remains relatively high. This resilience can be attributed to reduced quantum transmissions and the absence of global model broadcasting, which limits cumulative noise amplification across rounds. At $p = 10^{-4}$, Decentralized QFL achieves performance comparable to the noiseless case.

When Light-cone feature selection is combined with noisy communication (Fig.~\ref{fig:exp3_lightcone_rsna} and~\ref{fig:exp3_lightcone_kidney}), training becomes more unstable compared to Full aggregation. Since both parameter reduction and quantum channel noise introduce perturbations, their effects accumulate. This instability is pronounced in the RSNA dataset (Fig.~\ref{fig:exp3_lightcone_rsna}), and is further amplified in the Kidney CT-scan dataset (Fig.~\ref{fig:exp3_lightcone_kidney}), where accuracy remains low and exhibits high variance across noise levels even with Steane encoding. This may reflect the higher baseline accuracy of the Kidney task, which leaves less tolerance for the compounding perturbations introduced by parameter reduction and channel noise.

\subsubsection*{Quantum Error Correction and Noise Threshold}

We evaluate the effectiveness of quantum error correction using the Steane code. Without error correction, the effective noise threshold lies around $p \approx 10^{-4}$, beyond which learning degrades substantially. With Steane encoding, the tolerable noise range extends to $p \in [10^{-3}, 10^{-2}]$, representing an order-of-magnitude improvement in the noise threshold. As shown in Fig.~\ref{fig:exp3_centralized_rsna},~\ref{fig:exp3_centralized_kidney}, ~\ref{fig:exp3_decentralized_rsna}, and~\ref{fig:exp3_decentralized_kidney}, training with Steane encoding demonstrates stable convergence at noise levels where unencoded circuits fail entirely, confirming that error correction effectively mitigates channel noise. In contrast, the Light-cone configurations (Fig.~\ref{fig:exp3_lightcone_rsna} and~\ref{fig:exp3_lightcone_kidney}) show that Steane encoding provides limited benefit when combined with parameter reduction, suggesting that the compounding effect of parameter reduction and channel noise is not fully addressed by error correction alone.

However, Steane encoding requires seven physical qubits per logical qubit, increasing the quantum communication volume by a factor of 7. Under the current 6-qubit configuration, this raises the effective qubit count to 42 and the total transmission cost to $R_{\text{Steane}} = 7 \cdot 3\,NMP = 21\,NMP$ for Centralized QFL, giving a total of $R_{\text{Steane}} = 21\,TNMP = 630\,NMP$ over $T = 30$ rounds (a sevenfold increase compared to the unencoded total of $90\,NMP$). Although operational gate errors are not modeled in this simulation, practical implementations would additionally require encoding and decoding operations. Therefore, while error correction restores learning capability under heavy noise and extends the effective noise threshold by roughly one order of magnitude, practical deployment must carefully balance this mitigation benefit against the sevenfold increase in transmission overhead.

\section{Conclusion}
\label{sec:conclusion}

We implemented and systematically evaluated QFL for privacy-preserving medical image diagnosis, investigating two complementary strategies to reduce quantum transmissions: structured parameter reduction via Light-cone feature selection and architectural transformation through Hybrid QFL. The results demonstrated that each method reduces the quantum communication volume while maintaining comparable predictive performance and stable convergence, showing that communication-efficient QFL can be realized through principled architectural design rather than naive full-parameter aggregation.

We further showed that while depolarizing noise rapidly degrades learning beyond a certain threshold, the Steane error correction code restores stable convergence under high-noise conditions, confirming its potential effectiveness for practical deployment.

Overall, this work provides a concrete design framework for communication-efficient and noise-aware QFL, clarifying key trade-offs among convergence speed, communication cost, and noise resilience toward the practical realization of distributed quantum-enhanced machine learning systems. Unlike classical secure aggregation approaches, the proposed framework offers information-theoretic security guarantees that remain valid regardless of future advances in computational power, making it a promising foundation for privacy-preserving medical data utilization across institutions.

\appendices
\section{Train Dataset Distribution}
\label{sec:appendix}

Tables~\ref{table:rsna_distribution} and \ref{table:kidney_distribution} describe the distribution of training datasets among clients for the settings of $N=3,\, 5,\,7$.

This study adopts a quantity skew setting as the non-IID data distribution across clients. In this setting, class label ratios remain approximately uniform across clients, but the number of samples per client varies substantially. As shown in Table~\ref{table:rsna_distribution} and~\ref{table:kidney_distribution}, the sample size imbalance ratio reaches approximately 16:1 between the largest and smallest clients in the 7-client RSNA setting and the 7-client Kidney CT-scan setting.

While this represents a realistic and challenging form of heterogeneity reflecting institutional differences in data collection capacity, we acknowledge that other forms of non-IID data, including label distribution skew and feature distribution skew, are not evaluated in this study. Investigating the robustness of the proposed QFL framework under more diverse heterogeneity settings remains an important direction for future work.

\begin{table}[htbp]
\centering
\caption{RSNA Pneumonia Train Dataset Distribution per Clients}
\label{table:rsna_distribution}
\begin{tabularx}{0.36\textwidth}{@{} ccrrr @{}}
\toprule
\textbf{Setting} & \textbf{Client} & \textbf{Normal} & \textbf{Pneumonia} & \textbf{Total} \\
\midrule
\multirow{3}{*}{3 Clients}
  & A & 1,249 & 904   & 2,153 \\
  & B & 3,143 & 2,322 & 5,465 \\
  & C & 2,461 & 1,748 & 4,209 \\
\midrule
\multirow{5}{*}{5 Clients}
  & A &   918 &   657 & 1,575 \\
  & B & 2,297 & 1,701 & 3,998 \\
  & C & 1,790 & 1,288 & 3,078 \\
  & D & 1,467 & 1,050 & 2,517 \\
  & E &   381 &   278 &   659 \\
\midrule
\multirow{7}{*}{7 Clients}
  & A &   851 &   612 & 1,463 \\
  & B & 2,142 & 1,573 & 3,715 \\
  & C & 1,646 & 1,214 & 2,860 \\
  & D & 1,368 &   971 & 2,339 \\
  & E &   361 &   248 &   609 \\
  & F &   350 &   259 &   609 \\
  & G &   135 &    97 &   232 \\
\bottomrule
\end{tabularx}
\end{table}
 
\begin{table}[htbp]
\centering
\caption{Kidney CT-scan Train Dataset Distribution per Clients}
\label{table:kidney_distribution}
\begin{tabularx}{0.33\textwidth}{@{} ccrrr @{}}
\toprule
\textbf{Setting} & \textbf{Client} & \textbf{Normal} & \textbf{Disease} & \textbf{Total} \\
\midrule
\multirow{3}{*}{3 Clients}
  & A &   661 &   925 & 1,586 \\
  & B & 1,635 & 2,391 & 4,026 \\
  & C & 1,258 & 1,842 & 3,100 \\
\midrule
\multirow{5}{*}{5 Clients}
  & A &   482 &   678 & 1,160 \\
  & B & 1,224 & 1,721 & 2,945 \\
  & C &   898 & 1,369 & 2,267 \\
  & D &   726 & 1,128 & 1,854 \\
  & E &   224 &   262 &   486 \\
\midrule
\multirow{7}{*}{7 Clients}
  & A &   443 &   635 & 1,078 \\
  & B & 1,127 & 1,610 & 2,737 \\
  & C &   845 & 1,262 & 2,107 \\
  & D &   694 & 1,029 & 1,723 \\
  & E &   169 &   280 &   449 \\
  & F &   192 &   257 &   449 \\
  & G &    84 &    85 &   169 \\
\bottomrule
\end{tabularx}
\end{table}

\section*{Acknowledgments}

This work  was supported by JST Moonshot R\&D Grant Number JPMJMS226C. SK and TS are also supported by JST COI-NEXT Grant Number JPMJPF2221.

\section*{Author's Contribution}

S.K. conducted the primary research, analyzed the data, and wrote the main manuscript. H.K. conceived and designed the study, developed the analytical framework, provided medical insight regarding data interpretation and applicability, supervised the research activities, and critically revised the manuscript for important intellectual content. T.S. supervised the overall project. All authors reviewed and approved the final manuscript.

\bibliographystyle{IEEEtran}
\bibliography{kame}

\end{document}